\input{aipcheck}

\documentclass[
    ,final            
  ,cmfonts]
  {aipproc}
\usepackage{amsfonts,amssymb,amsmath}
\layoutstyle{6x9}

\newcommand{\be}{\begin{equation}}
\newcommand{\bea}{\begin{eqnarray}}
\newcommand{\ee}{\end{equation}}
\newcommand{\eea}{\end{eqnarray}}
\newcommand{\bpi}{\begin{picture}}
\newcommand{\bce}{\begin{center}}

\def\g{\widetilde{{\rm I}\hspace{-0.07cm}\Gamma}}

\def\gv{\widetilde{{\rm I}\hspace{-0.07cm}\Gamma}}
\newcommand{\D}{\displaystyle}

\begin{document}

\title{Effective gluon mass and \\ infrared fixed point in QCD}

\classification{
12.38.Lg, 
12.38.Aw  
}

\keywords{Schwinger-Dyson equations, pinch technique, gluon propagator. }

\author{Arlene~C.~Aguilar}{
  address={Departamento de F\'\i sica Te\'orica
and IFIC, Centro Mixto,\\ Universidad de Valencia -- CSIC \\
E-46100, Burjassot, Valencia, Spain}
}

\author{Joannis Papavassiliou}{
  address={Departamento de F\'\i sica Te\'orica
and IFIC, Centro Mixto,\\ Universidad de Valencia -- CSIC \\
E-46100, Burjassot, Valencia, Spain}
}

\begin{abstract}
We report on  a special type of solutions for  the gluon propagator of
pure QCD,  obtained from the  corresponding non-linear Schwinger-Dyson
equation  formulated in  the  Feynman gauge  of  the background  field
method.  These solutions reach a finite value in the deep infrared and
may be fitted using a massive propagator,  with the crucial
characteristic  that the  effective ``mass''  employed depends  on the
momentum  transfer.  Specifically,  the gluon  mass falls  off  as the
inverse square of the  momentum, as expected from the operator-product
expansion.   In addition,  one  may define  a dimensionless  quantity,
which constitutes  the generalization in a non-Abelian  context of the
universal QED effective charge.  This strong effective charge displays
asymptotic freedom in the ultraviolet whereas in the low-energy regime
it freezes at  a finite value, giving rise to  an infrared fixed point
for QCD.
\end{abstract}

\maketitle


\section{}
\vspace{-1.0cm}

A plethora of theoretical and phenomenological studies 
spanning more than two decades have 
corroborated the 
possibility of 
describing the infrared (IR) sector of QCD in terms of an effective gluon  mass
(for an extended list of references see~\cite{Aguilar:2006gr}).
According to this picture, 
even though the gluon is massless at the level of the fundamental Lagrangian,
and remains massless to all order in perturbation theory, 
the non-perturbative QCD dynamics generate 
an effective, momentum-dependent mass, without affecting 
the   local  $SU(3)_c$   invariance, which remains intact~\cite{Cornwall:1981zr}.

The most  standard way for  studying such a 
non-perturbative effect in the  continuum is  the
(appropriately truncated)
Schwinger-Dyson  equation   (SDE) for the gluon propagator $\Delta_{\mu\nu}(q)$,
defined (in the Feynman gauge) as 
\begin{equation}
\Delta_{\mu\nu}(q)= {-\D i}\left[{\rm P}_{\mu\nu}(q)\Delta(q^2) + 
\frac{q_{\mu}q_{\nu}}{q^4}\right]\,, \quad\quad\quad
{\rm P}_{\mu\nu}(q)= \ g_{\mu\nu} - \frac{\D q_\mu q_\nu}{\D q^2}\,.
\label{prop_cov}
\end{equation}
Specifically, one looks for 
  solutions having $\Delta(q^2)$ reaching finite
(non-vanishing) values in the  deep infrared,  
that may
be  fitted  by     ``massive''  propagators  of   the 
form $\Delta^{-1}(q^2)  =  q^2  +  m^2(q^2)$. 
The  crucial  characteristic is  that $m^2(q^2)$ is  not ``hard'', but
depends non-trivially  on the momentum  transfer $q^2$.  
When the  
renormalization-group 
logarithms are  properly taken into  account, one obtains  in addition
the  non-perturbative  generalization  of  $g^2(q^2)$, 
the  QCD  running  coupling (effective charge).
The presence of $m^2(q^2)$ in the argument of $g^2(q^2)$
tames  the   Landau  singularity   associated   with  the
perturbative $\beta$  function, and the resulting  effective charge is
asymptotically free in  the ultraviolet (UV), 
``freezing'' at a  finite value in the IR.

The  running   of  $m^2(q^2)$  is   of  central  importance   for  the
self-consistency of  this approach, mainly because the  value of
$\Delta^{-1}(0)$ is  determined by integrals  involving $\Delta(q^2)$,
$m^2(q^2)$,  and  $g^2(q^2)$  over  the entire  range  of  (Euclidean)
momenta. The  UV convergence of  these integrals depends  crucially on
how  $m^2(q^2)$ behaves  as  $q^2\to\infty$. If  $m^2(q^2)$ drops  off
asymptotically  faster  than  a  logarithm, then  $\Delta^{-1}(0)$  is
finite.  This,   in  turn,  is  crucial  because   the  finiteness  of
$\Delta^{-1}(0)$ guarantees essentially the renormalizability of QCD.

In earlier studies of {\it linear} SDE~\cite{Cornwall:1981zr,Aguilar:2006gr}
the $m^2(q^2)$ obtained drops in the 
deep UV as an inverse power of a logarithm. 
The main result reported in this talk is the existence of 
a new type of  solutions for $m^2(q^2)$ that drop asymptotically as 
an {\it inverse power} of momentum (multiplied by logarithms)~\cite{Aguilar:2007ie}.

\begin{figure}[h]
{\includegraphics[scale=1.0]{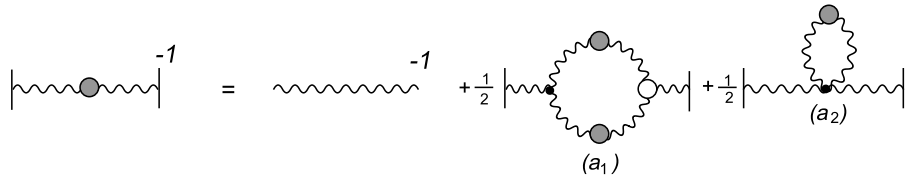}}
\caption{The gluonic ``one-loop dressed'' contributions to the SDE.}
\label{f1}
\end{figure}

These solutions are found in the study of {\it nonlinear} SDE,
in the framework defined from the combination of the 
Pinch Technique (PT)~\cite{Cornwall:1981zr,Cornwall:1989gv,Binosi:2002ft} and 
the Feynman gauge of the Background Field Method (BFM)~\cite{Abbott:1980hw,Sohn:1985em}, 
known as PT-BFM truncation scheme~\cite{Aguilar:2006gr}.
One of the  most powerful features of
the PT-BFM formalism is that, by virtue  of  the Abelian Ward identities  satisfied by   
the various vertices, gluonic   and  ghost   contributions   are  {\it
separately}   transverse,    within   {\it   each}    order   in   the
``dressed-loop'' expansion~\cite{Aguilar:2006gr}. This, in turn, 
allows one to truncate the series meaningfully, by considering only 
the  diagrams $({\bf a_1})$ and $({\bf a_2})$ shown in Fig.\ref{f1}, 
(no ghosts included), without compromising the transversality of the answer.

In order to reduce the algebraic complexity 
of the problem, we perform one additional approximation, dropping
the longitudinal terms  from the gluon propagators inside the integrals, i.e. we set
${\Delta}_{\alpha\beta} \to -ig_{\alpha\beta} {\Delta}$. Omitting these terms does not interfere with the
transversality of the resulting propagator, provided that one drops, 
at the same time, the  longitudinal pieces in the WI of Eq.(\ref{ward1}) \cite{Aguilar:2006gr,Aguilar:2007ie}.

After these steps, the scalar function, $\Delta^{-1}(q^2) = q^2 + i \Pi(q^2)$, (where $\Pi(q^2)$ is the gluon-self energy given by the diagrams  $({\bf a_1})$ and $({\bf a_2})$ in  Fig.\ref{f1} ) can be written as
\bea
i {\rm P}_{\mu\nu}(q) \Delta^{-1}(q^2)  
= i {\rm P}_{\mu\nu}(q) \,q^2\!\!\!\! &-& \!\!\!\!\frac{C_{\rm A} g^2}{2}\,
 \int\!  [dk]\,
\widetilde{\Gamma}_{\mu}^{\alpha\beta}
{\Delta}(k)
{\g}_{\nu\alpha\beta} 
{\Delta}(k+q) \nonumber \\
&+&\,4 \,C_{\rm A} g^2\, g_{\mu\nu}
\int\!  [dk] \, {\Delta}(k) \,,
\label{polar2}
\eea

where  the tree-level vertex $\widetilde{\Gamma}_{\mu\alpha\beta}$ appearing in $(\ref{polar2})$
is given by 
\be
\widetilde{\Gamma}_{\mu\alpha\beta}(q,p_1,p_2)=  
(p_1-p_2)_{\mu} g_{\alpha\beta} + 2q_{\beta}g_{\mu\alpha} - 2q_{\alpha}g_{\mu\beta} \,,
\label{gfey}
\ee
and ${\g}_{\nu\alpha'\beta'} $  represents the full three-gluon vertex.

As a next step we will employ 
the ``gauge technique"~\cite{Salam:1963sa}, expressing 
$\gv$ as a functional of $\Delta$, 
in such a way as to satisfy (by construction) the all-order Ward identity 
\begin{equation}
q^{\mu}\gv_{\mu\alpha\beta}(q,p_1,p_2) = 
i[{\Delta}^{-1}_{\alpha\beta}(p_1) - {\Delta}^{-1}_{\alpha\beta}(p_2)]\,,
\label{ward1}
\end{equation}
characteristic of the PT-BFM.
Specifically, we propose the following form for the vertex~\cite{Aguilar:2007ie}
\be
\gv^{\mu\alpha\beta}= 
L^{\mu\alpha\beta} + T_1^{\mu\alpha\beta} + T_2^{\mu\alpha\beta}\,,
\label{gtvertex}
\ee
with 
\bea
L^{\mu\alpha\beta}(q,p_1,p_2)  &=&  \widetilde{\Gamma}^{\mu\alpha\beta}(q,p_1,p_2)
+i g^{\alpha\beta}\, \frac{q^{\mu}}{q^2}\,
\left[ {\Pi}(p_2) - {\Pi}(p_1)\right]\,,
\nonumber\\
T_1^{\mu\alpha\beta}(q,p_1,p_2) &=& 
- i\frac{c_1}{q^2}\left(q^{\beta}g^{\mu\alpha} - q^{\alpha}g^{\mu\beta}\right)
\left[{\Pi}(p_1)+ {\Pi}(p_2)\right]\,,\nonumber\\
T_2^{\mu\alpha\beta}(q,p_1,p_2) &=& -i c_2 \left(q^{\beta}g^{\mu\alpha} - q^{\alpha}g^{\mu\beta}\right)
\left[\frac{{\Pi}(p_1)}{p_1^2}+ \frac{{\Pi}(p_2)}{p_2^2}\right]\,.
\label{LT1T2}
\eea

Then, substituting Eqs.(\ref{gfey}) and (\ref{gtvertex}) into (\ref{polar2}), introducing
$q^2 \equiv x$, $k^2 \equiv y$, and defining the renormalization-group
 invariant quantity ${d}(q^2) = g^2 \Delta(q^2)$ , we arrive at
\be
d^{-1}(x) = K^{\prime}x + \tilde{b} \sum_{i=1}^8 {\widehat A}_i(x)\, + d^{-1}(0)\,,
\label{rgisde}
\ee
with
\bea
{\widehat A}_1(x) &=& - \left(1+\frac{6 c_2}{5}\right) x \int_{x}^{\infty}\!\!\! dy \,y\, {\cal L}^{\,2}(y) d^{\,2}(y)\,,   
\nonumber\\ 
{\widehat A}_2(x) &=& \frac{6 c_2}{5} x \int_{x}^{\infty}\!\!\! dy\, {\cal L}(y) d(y)\,, 
\nonumber\\
{\widehat A}_3(x) &=& -\left(1+\frac{6 c_2}{5}  \right) x\, {\cal L}(x) d(x)\int_{0}^{x}\!\!\! dy\, y\, {\cal L}(y) d(y)\,, 
\nonumber\\ 
{\widehat A}_4(x) &=& \left(-\frac{1}{10} -  \frac{3c_2}{5} + \frac{3c_1}{5}\right)
\int_{0}^{x}\!\!\! dy\, y^2\, {\cal L}^{\,2}(y) d^{\,2}(y)\,, 
\nonumber \\
{\widehat A}_5(x) &=& - \frac{6}{5}\bigg(1+ c_1\bigg)  {\cal L}(x) d(x)\int_{0}^{x}\!\!\! dy \,y^2\,{\cal L}(y) d(y)\,, 
\nonumber \\
{\widehat A}_6(x) &=& \frac{6c_2 }{5} \int_{0}^{x}\!\!\! dy\, y \,{\cal L}(y) d(y)\,,
\nonumber\\
{\widehat A}_7(x) &=& \frac{2}{5}\, {\cal L}(x)\,\frac{d(x)}{x}\int_{0}^{x}\!\!\! dy\, y^3\,  {\cal L}(y) d(y)\,, 
\nonumber\\
{\widehat A}_8(x) &=& \frac{1}{5x} \int_{0}^{x} \!\!\! dy\, y^3\,  {\cal L}^{\,2}(y) d^{\,2}(y)\,.
\label{hatAi}
\eea
The renormalization constant $ K^{\prime}$ is fixed by the condition $d^{-1}(\mu^2)=\mu^2/g^2$, (with 
$\mu^2\gg \Lambda^2$), and 
${\cal L}(q^2) \equiv \tilde{b}\ln\left(q^2/\Lambda^2\right)$, 
where $\Lambda$ is QCD mass scale.
Due to the poles contained in the Ansatz for $\gv$, $d^{\,-1}(0)$ does not vanish, 
and is given by the (divergent) expression
\be
d^{\,-1}(0) = \frac{3 \tilde{b}}{5\pi^2} 
\Bigg[2(1+c_1)\int  d^4 k \,{\cal L}(k^2)\, d(k^2) - (1+2c_1)\int\,d^4 k \, k^2 \,{\cal L}^{\,2}(k^2)\,d^{\,2}(k^2) \Bigg]\,,
\label{D0}
\ee
which can be made finite using dimensional regularization, and assuming 
that $m^2(q^2)$ drops sufficiently fast in the UV~\cite{Aguilar:2006gr}.

 In order to determine the asymptotic behavior that Eq.(\ref{rgisde}) predicts for $m^2(x)$ at large $x$, we perform the following replacements in the r.h.s. of (\ref{hatAi})
\be
{x{\cal L}(x)d(x)\to 1}, \quad {\cal L}(x)d(x)\to 1/x, \quad 
{\cal L}(y)d(y) = \tilde\Delta(y), \quad \tilde\Delta(y) =  \frac{1}{y + m^2(y)}\,.
\ee

Next, use the identity
\mbox{$y\tilde\Delta(y) = 1 - m^2(y)\tilde\Delta(y)$} in all ${\widehat A}_i(x)$, 
keeping only terms linear in $m^2$ . Then separate all contributions that go like $x$ from those that go like 
$m^2$ on both sides, and match them up~\cite{Cornwall:1985bg}. This gives rise to two independent 
equations, one for the ``kinetic'' term, which simply  
reproduces the asymptotic behavior $x\ln x$ on both sides, 
and an equation for the terms with $m^2(x)$, given by
\bea
m^2(x)\ln x  &=& {\cal C}
-a_1 \int_{x}^{\infty} dy \,m^2(y) \tilde\Delta(y) 
+ \frac{a_2}{x} \int_{0}^{x} dy \,y m^2(y) \tilde\Delta(y)\,\, \nonumber\\
&&+  \frac{a_3}{x^2}\int_{0}^{x} dy \,y^2 m^2(y) \tilde\Delta(y)
+ a_4 x \int_{x}^{\infty} dy\, m^2(y) \tilde\Delta^2(y)\,,
\label{meq2}
\eea
with
\be
a_1 = \frac{6}{5} (1+c_2-c_1) \,,\,\,\,\,\,\,
a_2 = \frac{4}{5} + \frac{6c_1}{5} \,,\,\,\,\,\,\,
a_3 = - \frac{2}{5}\,, \,\,\,\,\,\,
a_4 = 1+ \frac{6c_2}{5} \,,
\label{ai}
\ee
and
\be
{\cal C} \equiv  \tilde{b}^{-1} d^{-1}(0) + a_1 \int_{0}^{\infty} dy \,m^2(y) \tilde\Delta(y)\,.
\label{C0}
\ee
Now,  the  important  point  to  appreciate  is  that,  in  order  for
(\ref{meq2}) to have solutions vanishing in the UV, it is necessary to
be sure  that the  constant term on  the r.h.s. vanishes,  i.e. ${\cal
C}=0$. Since  we know that  $d^{-1}(0)$ and the integral  appearing in
the  r.h.s. of  Eq.(\ref{C0}) are  manifestly positive  quantities, it
follows immediately  that the ${\cal  C}$ will be  zero if and  only if
$a_1<0$.  Notice that  Eq.(\ref{C0})  restricts the  range of  allowed
values  of the parameters  $c_1$ and  $c_2$ through  Eq.(\ref{ai}). In
addition, and more importantly,  it constrains the momentum dependence
of $m^2(x)$  in the  IR and intermediate regimes to  be such
that  both terms  on the  r.h.s of  Eq.(\ref{C0}) cancel  against each
other.

Assuming that Eq.(\ref{C0}) is satisfied,  it can be shown that Eq.(\ref{meq2})
admits    the   following    asymptotic    solutions   for    $m^2(x)$
\cite{Aguilar:2007ie},
\be
m_1^2(x)  =  \lambda_1^2 (\ln x)^{-(1+\gamma_1)} \,, \qquad
m_2^2(x)  = \frac{\lambda_2^4}{x} (\ln x)^{\gamma_2-1}\,, 
\label{m2}
\ee
where $\lambda_1$ and $\lambda_2$ are two mass-scales,
and $\gamma_1=-a_1$, $\gamma_2 = a_2$.

The first type of solutions, $m_1^2(x)$, are familiar from studying 
linearized versions of Eq.(\ref{polar2}), see for example
\cite{Cornwall:1981zr,Aguilar:2006gr}. 
The second type of solutions, $m_2^2(x)$, displaying power-law running,
are particularly interesting because they are derived for the first time in 
the context of SDE. The possibility of an effective gluon mass dropping in the UV as an 
{\it inverse power of  the momentum} was first 
conjectured in~\cite{Cornwall:1981zr}, and was subsequently obtained in the context of the  
operator-product expansion ~\cite{Lavelle:1991ve}; there the resulting gluon self-energy was
identified as the effective gluon  mass, leading to the relation 
\mbox{$m^2(x)\sim  \langle G^2 \rangle/x$}, where $\langle G^2 \rangle$ is the dimension four 
gauge-invariant gluon condensate.

\begin{figure}[ht]
\hspace{-1.5cm}
\includegraphics[scale=2.0]{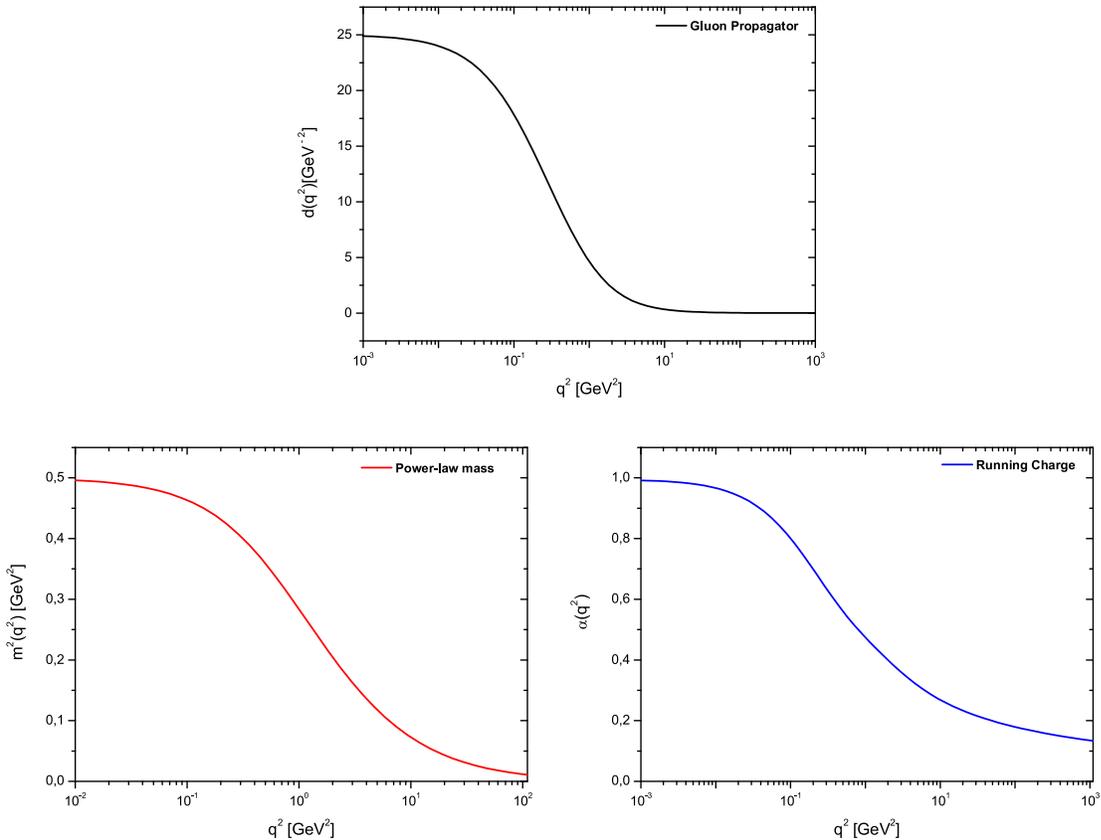}
\vspace{-0.5cm}
\caption{\small{Upper panel: the numerical solution for $d(q^2)$. Lower panels: 
Left: dynamical mass with power-law running, for 
$m_0^2=0.5    \;\mbox{GeV}^{\,2}$     and    $\rho=1.046$    in  
Eq.(\ref{dmass_power}). Right: the running charge,
$\alpha(q^2)=g^2(q^2)/4\pi$,  fitted  by Eqs.(\ref{GNP}) and (\ref{func_fit}).}}
\label{fc}
\end{figure}

Which of the two types 
of solution will be actually realized depends on  the   details  of  the
three-gluon vertex, $\g$, and more specifically on the values of the  parameters $c_1$ and
$c_2$. Our numerical analysis reveals that the sets of
values  for $c_1$  and $c_2$  giving rise  to logarithmic  running
belong to  an interval that  is disjoint and well-separated  from that
producing power-law running.
In what follows we will focus our attention on the latter type of solutions. In
Fig.\ref{fc} we present typical solutions for the $d(q^2)$, $m^2(q^2)$ and 
the effective charge  $\alpha(q^2)=g^2(q^2)/4\pi$. 

\vspace{0.5cm}

The way to extract from  $d(q^2)$ the corresponding $m^2(q^2)$ and $g^2(q^2)$
is by casting the numerical solutions shown in Fig.\ref{fc} into the form 
\be
d(q^2) = \frac{g^2(q^2)}{q^2+m^2(q^2)} \,, \quad
g^2(q^2) = \bigg[ \tilde{b}\ln\left(\frac{q^2 + f(q^2, m^2(q^2))}{\Lambda^2}\right)\bigg]^{-1}\,,
\label{GNP}
\ee
with
\be
f(q^2, m^2(q^2)) = \rho_{\,1} m^2(q^2)+ \rho_{\,2} \frac{m^4(q^2)}{q^2+m^2(q^2)} 
+\rho_{\,3} \frac{m^6(q^2)}{[q^2+m^2(q^2)]^{\,2}} \,,
\label{func_fit}
\ee
where $\rho_{\,1}$, $\rho_{\,2}$, and  $\rho_{\,3}$ are fitting parameters.

The functional form  used for the running mass is
\be
m^2(q^2)=\frac{m^4_0}{q^2+m^2_0}\Bigg[\ln
\left(\frac{q^2+\rho\,m^2_0}{\Lambda^2}\right)\Big/\ln\left(\frac{\rho\,m^2_0}{\Lambda^2}\right) \Bigg]^{\gamma_2-1} \,.
\label{dmass_power}
\ee 
In the deep UV Eq.(\ref{dmass_power}) goes over to $m^2_2(q^2)$, whereas 
at $q^2=0$ it reaches the finite value $m^2(0)=m^2_0$.
Note that $ f(q^2, m^2(q^2))$ is such that $ f(0, m^2(0)) >0$; as a result
the perturbative Landau  pole in  the  running coupling is tamed, and
$g^2(q^2)$ reaches a  finite positive value  at $q^2=0$,
leading to an {\it infrared fixed point}
\cite{Cornwall:1981zr,Aguilar:2002tc,Brodsky:2007wi}.

To summarize  our results,  from a gauge-invariant  SDE for  the gluon
propagator  we have derived  an integral  equation that  describes the
running of the  effective gluon mass in the  UV, and have demonstrated
that, depending on the values of two basic parameters appearing in the
three-gluon vertex, one finds solutions that drop as inverse powers of
a  logarithm  of  $q^2$,  or  much  faster, as  an  inverse  power  of
$q^2$. Moreover,  we have  extracted an asymptotically  free effective
(running) charge,  that freezes  in the low-momentum  region, implying
the existence of a IR fixed point for QCD.


\vspace{-0.5cm}
\begin{theacknowledgments}

This work was supported by the Spanish MEC under the grants FPA 2005-01678 and 
FPA 2005-00711.  
The research of JP is funded by the Fundaci\'on General of the UV. 

\end{theacknowledgments}



\bibliographystyle{aipproc}   





\end{document}